\documentstyle[12pt,psfig]{article}
\def\reference{\parskip 0pt\par\noindent\hangindent 0.5 truecm}

\begin{document}
%
%
\title{Local HI:  Constraints on
the evolution of the HI content of the Universe}
\author{F. H. Briggs\\Kapteyn Astronomical Institute\\fbriggs@astro.rug.nl  }
\date{} 
\maketitle

\begin{abstract}
Analyses of QSO absorption lines are showing that HI content has
evolved over the redshift range z=5 to z= 0.  The 21cm line measurements
of the z=0 HI content avoid several biases inherent in the
absorption line technique, such as the influence of evolving 
dust content in the absorbers, and will produce a reliable measure
to anchor theories of galaxy evolution.
Examples of important questions to
be addressed by local HI surveys
are: (1) is there a significant population of gas-rich galaxies 
or intergalactic clouds that are missing from the census of optically
selected galaxies?  (2) is there an adequate reservoir of neutral gas
to substantially 
prolong star formation at its present rate? and (3) are there massive
objects of such low HI column density that they can have escaped detection in 
the ``unbiased'' HI surveys that have been conducted so far?
\end{abstract}

\bigskip

\section{Evolution of the HI Content}

\begin{figure}
\centerline{
\psfig{figure=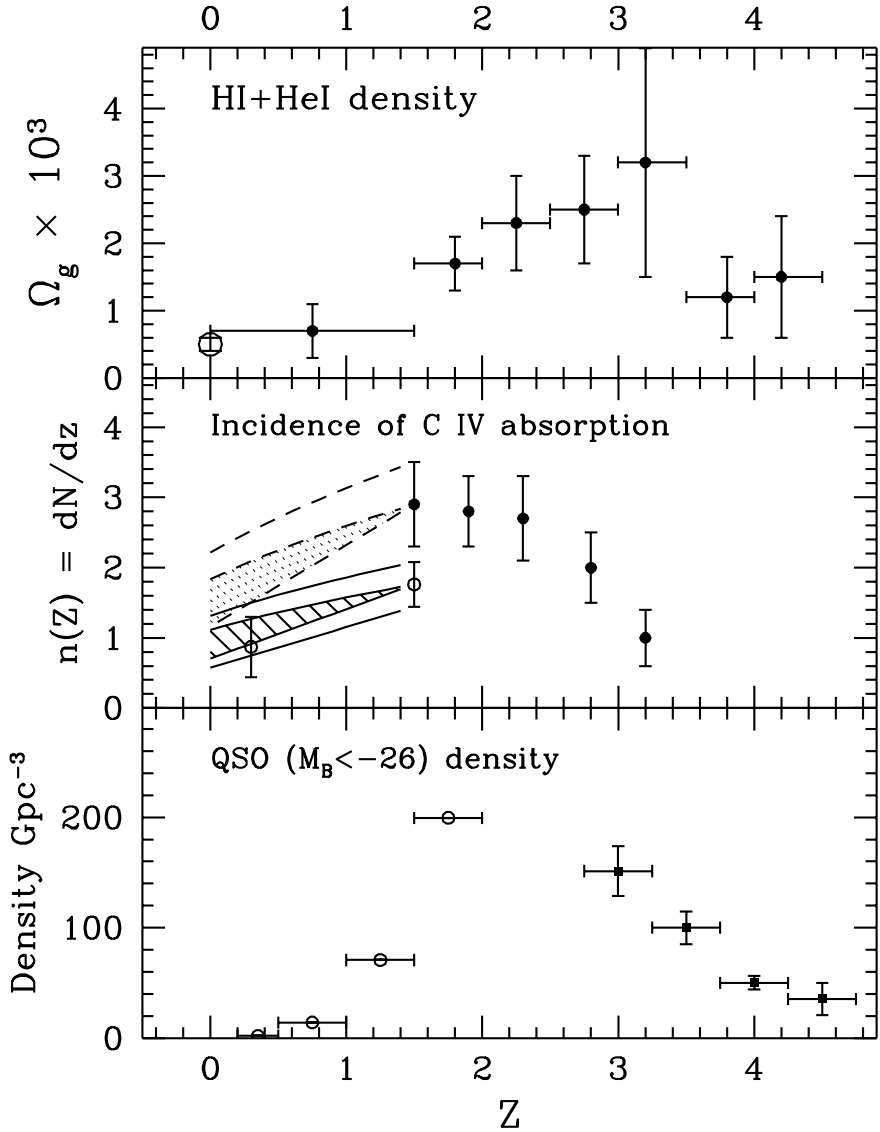,width=16cm}
}
\caption{Cosmological density of neutral gas, incidence 
of CIV absorption, and  comoving density of luminous QSOs
as a function
of redshift from QSO absorption-line statistics. 
{\it Top panel.}  Mean cosmological density of neutral gas,
$\Omega_g$, normalized to the critical density (Storrie-Lombardi et al
1996; Rao et al 1995 ($z=0$); Storrie-Lombardi \& Wolfe priv. comm.)
{\it Bottom panel.} Number of CIV metal-line absorption systems per unit redshift, $n(z)$ (Steidel 1990); $z=0.3$ point from
Bahcall et al 1993). Filled points
from Steidel indicate rest frame equivalent widths 
$W_{rest}({\lambda}1548) > 0.15$ \AA; open points are for 
$W_{rest}({\lambda}1548) >0.3 $ \AA. 
Hatched areas indicate the range ($0<q_o<1/2$) for unevolving
cross sections since $z=1.5$, beyond which redshift CIV can be
measured with ground-based telescopes. 
{\it Bottom panel.} Comoving density of optically selected QSOs:
filled squares from Schmidt et al 1994; open circles from Hewitt et al 1993).
$H_o = 50 $km~s$^{-1}$Mpc$^{-1}$, $q_o =1/2$}
\end{figure}

A measurement of the total HI content contained in nearby galaxies and
HI clouds is an important constraint in the bigger picture of galaxy
evolution on cosmological time scales. Although the neutral
gas in large galaxies at
present is often considered to be a minor component that is used
as tracer for kinematics or as a dwindling source of fuel for star
formation, there is now strong evidence from the studies of QSO 
absorption lines that the present HI content of the universe is but a fraction
of what it was at redshift $z\approx 3$. This finding comes from the
statistics of the ``damped Lyman-$\alpha$'' class of absorption line that
is identified with dynamically cold layers akin to the disks of familiar
nearby spiral galaxies (Wolfe et al 1986, Lanzetta et al 1995, Storrie-Lombardi
et al 1996).  Figure 1 summarizes the presently available measurements
of the neutral gas content
as a function of $z$  (Storrie-Lombardi et al 1996), 
along side a plot of the
incidence of CIV absorption lines, which provide an indication of the
cross section presented by clouds of ionized, metal-rich gas
(Steidel 1990, Bahcall et al 1993), and the
comoving density of luminous optically selected QSOs
(Schmidt et al 1994, Hewett et al 1993).   Taken 
literally, the evidence points to an epoch around $z\approx 3$ when
the neutral gas mass density, $\Omega_g$,
hit a maximum, at roughly the same time that the
CIV cross section for strong absorption lines 
($W_{rest}({\lambda}1548) > 0.15$ \AA)
began a sharp increase.  This is also a time when luminous 
QSOs were most abundant.  These indicators testify that we are seeing
substantial redistribution of gas, as witnessed by the formation of ionized
metal-rich galaxy halos and the efficient fueling of active galactic nuclei.
The surge in neutral gas content indicates that protogalactic
gravitational potentials
were deep enough that gas was confined to sufficiently high  density that
it was at least momentarily immune to ionization by star bursts and 
ionizing background radiation.  This may be the epoch that disk galaxies
formed as secondary infall of gas occured into galactic potentials formed
in the first round of galactic bulge formation. The disk formation would
be accompanied by halo enrichment, either by in situ star formation or
by metal pollution of the extended halo region by winds from the new
star forming regions of the disk. An alternative view is that
this is also likely to be an epoch when 
small protogalactic lumps are merging vigorously, and star burst within
the lumps would be effective at ejecting metals into an extended
region that would at later times constitute the ``halo'' region of
the merger product.

Figure 1 summarizes only the neutral atomic and ionized gas components. 
A complete balance requires an accounting for
all the universe's baryons, as gas is exchanged between neutral, ionized,
and molecular phases, as well as the path of stellar evolution
leading to the current state where far
more baryons are contained in stars than in neutral gas.  Although the present
HI content is only ${\sim}10$\% of the mass in stars, there was a period
at $z\approx 2.5$ when the HI content apparently was of order half of the
present stellar mass. Estimates of the mass content in ionized halos suggest
that they probably contain about ten times the mass contained in the
damped Lyman-$\alpha$ absorbers at their peak
(Petitjean et al 1993). This interpretation of the
CIV data relies on theoretical modeling, with large uncertainties due to
ionization level and carbon abundance. It is striking that the recent
HST observations (Bahcall et al 1993)
are consistent with no evolution in the absorption cross section
presented by high column density CIV systems from $z\approx 1.3$
(Steidel 1990) to $z\approx 0.3$, implying that large quantities
of ionized gas may still be present, either in the form of extended halos or
in intergalactic clouds whose mass could far exceed the visible
stellar mass in galaxies. The hydrogen neutral fraction of these
ionized clouds would
provide column densities of H$^0$ well below
the regime ordinarily probed by 21cm line observations. At present,
the integral molecular gas mass content of galaxiers
appears to be roughly equal to the
neutral atomic mass (cf. Kennicutt et al 1994).

\begin{figure}
\centerline{
\psfig{figure=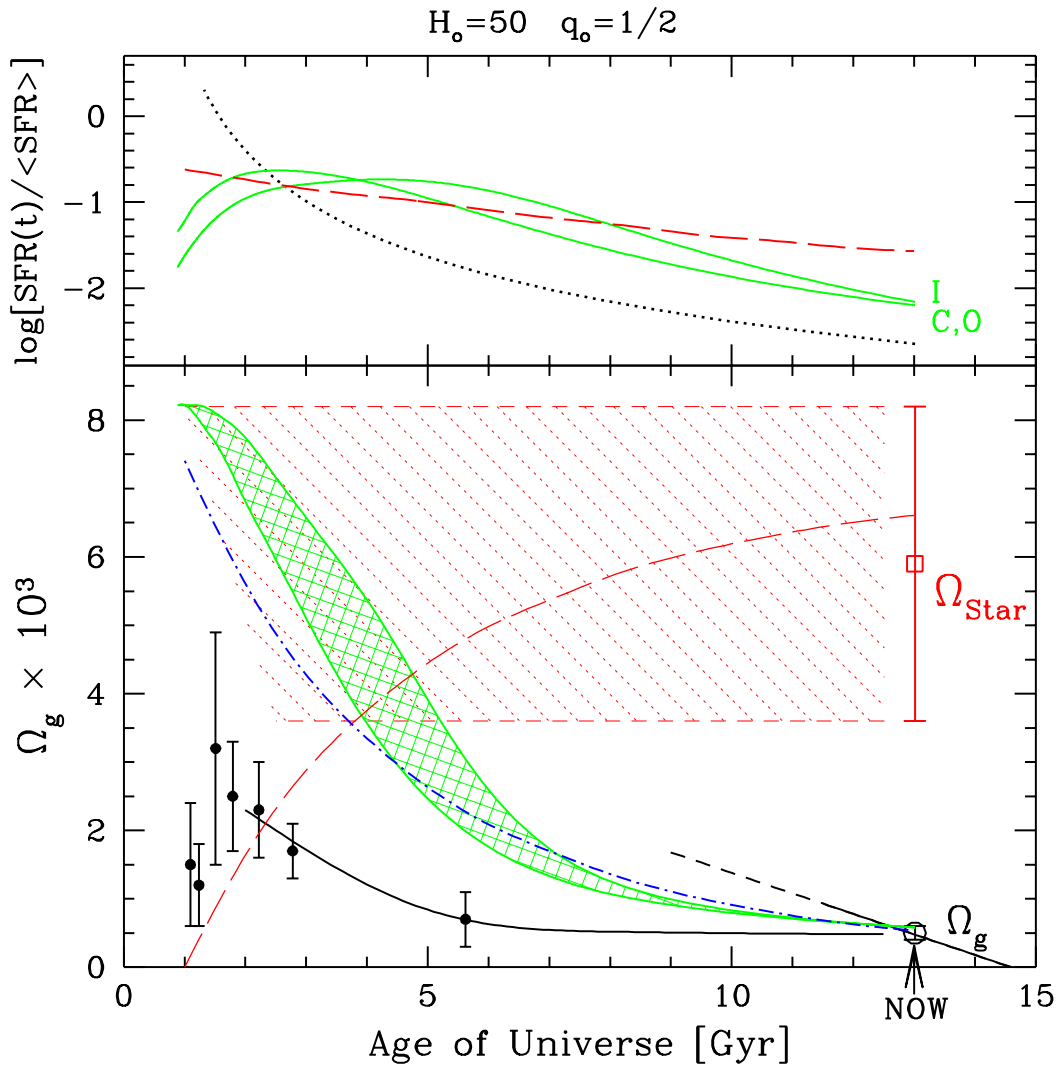,width=15cm}
}
\caption{Density of neutral gas as a function
of time compared with $\Omega_{stars}(z=0)$. $\Omega_g(z)$ from references
in Figure 1;  $\Omega_{stars}(z=0)$ from Lanzetta et al 1995. Rising
dashed curve is KTC model for increasing stellar mass; dot-dash is
KTC model for declining cold gas content, adjusted as described in text to
indicate only the atomic fraction.  The cross hatched band indicates the
range of models proposed by Pei and Fall for the true $\Omega_g(z)$, corrected
for selection effects caused by dusty damped Lyman-$\alpha$ absorbers.
}
\end{figure}

Surveys of nearby volumes in the 21 cm line are important in anchoring
the $z=0$ point of Figure 1.  A complete inventory of neutral
gas in the nearby Universe, of the sort that is being provided
by unbiased 21cm line surveys, will have tight error bars and thus
will carry large statistical weight in models that describe the
evolution of $\Omega_g$.  Note that the statistical errors of the
$z>1.6$ measurements of $\Omega_g$ are so large that these high $z$ points
are consistent with no evolution at all.  
In principle, the low $z$ regime ($0 < z < 1.6$) of the
diagram can be measured using QSO absorption line methods in much
the same way as the high $z$ points. However, the observations are 
difficult since the Lyman-$\alpha$ line is not redshifted into the
optical window until $z \geq 1.6$, so the low $z$ absorption line
work must be done
from space observatories such as IUE and HST (Lanzetta et al 1995,
Rao et al 1995).  

When the data points are plotted as a function of
time, as in Figure 2, it is clear that the single low $z$ QSO absorption
line point applies to a time span well longer than half the age of the Universe.
Several selection effects make it difficult to obtain a reliable
sampling of damped Lyman-$\alpha$ absorbers at low $z$.
Cosmological factors, as well as the apparent shrinking
of damped Lyman-$\alpha$ absorption cross section, $\sigma(t)$, with 
increasing age of the Universe (Lanzetta et al 1995),
act to make them very rare at
recent times: $dN/dt \propto \sigma(t)t^{-k}$, where $k=2$ for 
$q_o=1/2$ and $k=3$ for $q_o=0$.
There are also selection
effects that are likely to 
influence the QSO absorption line measurements and that
affect the low $z$ measurement most strongly: (1) The presence of
dust in disk galaxies is expected to become increasingly important
as they evolve and may act to selectively attenuate the light from
QSOs behind them, causing these lines of sight to be under represented
in QSO samples (Fall \& Pei 1989, Pei \& Fall 1995, Webster et al 1995)
although this view is contested (Boyle \& Di Matteo 1995). 
(2) Gravitational lensing
may act to selectively amplify background objects into QSO samples,
but also may bend the light path so as to dodge the high HI column
densities of the disk (Bartelmann \& Loeb 1996, Smette et al 1995a,1996). 

Figure 2 includes curves to indicate trends in stellar
evolution relative to the decline in $\Omega_g$ with time. Recent
analyses of the depletion of the integral mass content of
galaxies over cosmological times due to star formation
have been presented by Lanzetta et al (1995)
and Pei and Fall (1996).  A related study by
Kennicutt et al (1994, KTC)  addresses
the prolonging of the current star formation rate in $z\approx 0$ disks
due to delayed gas return as stellar populations age.
An example of the KTC models is presented in Figure 2 to illustrate both the
rise of stellar mass with time and decline of neutral gas content with time
for a disk system without added inflow or allowing mass to escape. 
For this display, 
the KTC model has been scaled so that the final stellar mass and
the final HI mass are consistent with the observations at $z\approx 0$;
the relative proportion of HI to H$_2$ has been adjusted for this
display to vary linearly with time
from $\rho_{HI}/(\rho_{HI}+\rho_{H_2}) = 1$ at high $z$ to 2/3 at
the present time.  
The slope of the KTC model at the time marked ``Now'' in Figure 2
can be compared with the steeper slope drawn to indicate the rate at
which the current star formation rate would consume the present atomic
hydrogen content, exhausting the supply in only $\sim$1.5 Gyr if there
were no additional reservoirs of molecular gas or contribution from
delayed stellar return.

The models of Lanzetta et al require a balance
between the decline of
$\Omega_g$ and a rise in stellar mass;  along with assumptions
that stellar return occurs instantaneously as each generation of 
stars is formed, this 
constrains the mean star formation rate history
of the Universe.  A consequence is that their models 
result in an uncomfortably large number of stars
at $z\approx 0$ with low metal abundances.  Pei and Fall suggest that
this problem can be solved with a family of dusty models, which imply
that the damped Lyman-$\alpha$ statistics drastically underestimate the
integral gas content at all redshifts.  The current generation of radio
telescopes are not sensitive enough to resolve this controversy by 
simply looking back to measure the integral HI content at $z\approx$ 1/2.
On the other hand, choosing complete samples of 
radio selected high $z$ quasars for background probes would remove
possible selection effects by dust.
 
At $z=0$, recent 21 cm line measurements are indicating
that the bulk of the atomic hydrogen content of the nearby
universe is bound into galaxies
with optical counterparts (Zwaan et al 1996, Schneider 1996, Szomoru
et al 1994, Henning 1995, Briggs 1990).  Furthermore the normalization
of seems to be well understood (cf. Zwaan et al 1996, Rao \& Briggs 1993),
although there is still concern over the normalization 
of even the optically determined 
luminosity function (cf Ellis et al 1996, Glazebrook et al 1995).  
Clearly the determination of the integral HI content 
is a measurement that the Parkes Multibeam Survey will clarify
since it will be complete, unbiased by extinction and optical surface
brightness, and will have well understood sensitivity limits.

\section{Distribution of $N(H^o)$ Column Densities}

Another place where future surveys can play
an important role will be in exploring lower HI column densities than
have been observed in the 21 cm line in the past.
QSO absorption line statistics over a wide range of Lyman-$\alpha$
line strengths specify that the incidence of absorption becomes
increasingly prevalent toward lower column densities, so that
along a randomly chosen sight line, the probability of interception
rises by roughly a factor of 10 for every decrease by a factor of 100
in $N_{HI}$.  This behavior is quantified by the $f(N_{HI})$
distribution, defined so that $f(N_{HI})dXdN_{HI}$ is the number of lines
detected within a range $dN_{HI}$ centered on $N_{HI}$ over a 
``normalized absorption distance'' $dX$, 
where $X= \frac{1}{2}[(1+z)^2-1]$ is the normalized absorption distance from
zero redshift to $z$ for $q_o=0$. 
The function $f(N_{HI})$ is roughly proportional to
$N^{-1.5}$ over the range $N_{HI} = 10^{13}$ to $10^{22}$ cm$^{-2}$, 
although there is evidence for subtle structure 
possibly related to  opacity in the Lyman continuum that occurs
for layers with
$N_{HI} > 3{\times}10^{17}$ cm$^{-2}$ (Petitjean et al 1993). At high
redshift, the frequency of absorption for these optically thick absorbers
is ${\sim}0.9$ per unit $X$ (Petitjean et al 1993); for 
$N_{HI} > 10^{18}$ cm$^{-2}$,
the frequency is roughly halved.  The derivation of the $f(N_{HI})$
at these $N_{HI}$ is especially uncertain, since the entire Lyman
series is heavily saturated in the regime where the Lyman continuum is
optically thick, and the damping wings, which permit an unambiguous
measure of column density, do not become readily observable
until $N_{HI}$ is well in excess of 10$^{20}$ cm$^{-2}$. Thus, column
density measurements in this regime are very uncertain, leading Petitjean et al
to plot only one point on their $f(N_{HI})$ diagram for $10^{17.7}$
to $10^{20.5 }$ cm$^{-2}$.

Recent large HI surveys with filled aperture telescopes
are routinely capable of detecting column densities below $10^{19}$ cm$^{-2}$
(Schneider 1996, Zwaan et al 1996, Briggs et al 1996), provided
the emission fills the telescope beam.  The Arecibo survey
by Sorar (1994; see also Briggs \& Sorar 1996) was optimized to be sensitive to
$N_{HI} = 10^{18}$ cm$^{-2}$ ($5\sigma$), and the survey
observed over 5000 independent beam areas
to a depth of 7500 km~s$^{-1}$, covering a total absorption path 
$\Delta X \approx 120$. The Arecibo beam subtends ${\sim}3$~Kpc at 3~Mpc
and 70~Kpc at 75~Mpc, which is a reasonable match to the cloud sizes
deduced for the Lyman-$\alpha$ forest (cf. Smette et al 1995b, 1992).
To date, only one of the 61 detections (Zwaan et al 1996)
has not been identified with a
high column density layer of the type associated with the neutral
intergalactic medium of a galaxy, implying that a separate population
of low column density objects can add only a small fraction
of the current integral HI content already identified with galaxies.

The high redshift $f(N_{HI})$
distribution would imply of order 50 interceptions in the range 
$10^{18}$ to $10^{19}$ cm$^{-2}$ for the pathlength $X$ explored
by the Arecibo survey.  
Where are they? At least a part of the discrepancy is likely due
to evolution of Lyman-$\alpha$ forest cloud population. An additional
observational problem is that the high
column density end of the forest cloud distribution
(around 10$^{17}$ cm$^{-2}$ and above) has
associated metal lines, such as CIV and MgII, which has historically
caused then to be identified with hypothetical galaxy halos;  single
dish observations seldom have the resolution to reliably
separate the HI signal from a halo of a spiral
galaxy from the bright signal originating in the main body of the
galaxy, unless there are strong kinematic effects that
create a difference in gas velocity as a function of radius and
the halo gas is very extended.

Further considerations in the study of this intermediate column density
range are the theoretical models that consider ionization of extended
gas around galaxies by the extragalactic ionizing background (Sunyaev
1969, Corbelli et al 1989, Maloney 1993, Charlton et al 1994).  Many of
these models predict a strong dip in $f(N_{HI})$ between 10$^{17.5}$ and
10$^{19.5}$ cm$^{-2}$.  If similar arguments apply to a population
of intergalactic clouds or super-LSB galaxies, then
an interesting experiment  now coming into the realm 
of possibility will be to push the sensitivity limits of the local
HI emission observations down to ${\sim}10^{17}$ cm$^{-2}$ where ionized layers
of high column density might be detected. 

\section{Where to find ``HI Clouds'' Now}

Some very interesting examples of HI without coincident
optical emission have turned up by chance in ``off-scans'' from
21cm line studies (Schneider 1989, Giovanelli \& Haynes 1989, 
Chengalur et al 1995, Giovanelli et al 1995), although all of these detections
appear to be associated in some way with nearby visible galaxies.  For
example, the 
protogalaxy of Giovanelli \& Haynes (1989) now appears in VLA
observations to more
closely resemble a galaxy with tidal remnant (Chengalur et al 1995)
than a single large cloud. An interesting consequence of high
sensitivity VLA surveys of HII galaxies and LSB dwarfs
(Taylor et al 1993; Taylor et al 1996) is an apparent 
increased probability of finding dim HI-rich
companions to the HII galaxies as compared with the LSBs.  This
kind of study will not be possible with the crude resolution of
a single-dish surveys.  The tendency of small galaxies to be found
in the vicinity of large ones rather than in isolated regions
(cf. Szomoru et al 1996) will complicate the derivation
of a HI-mass function that extends to faint masses when survey observations
are made with a large beam.  On the other hand, isolated HI
clouds (if they exist) and HI clouds associated with early type galaxies,
such as polar rings and fresh examples of the Leo Ring 
(Schneider 1989), should be readily
identified if confusion with other galaxies in the same field is
not too much of a problem.
 
\section{Conclusion}

Large sky surveys in the 21cm line will improve the
measurement of the integral HI content of the Universe, and this value
will find immediate use in anchoring theories of galaxy evolution. 
Telescopes with large beams will suffer from confusion, which
will complicate the determination
of the faint end slope of the HI-mass function. The low column
density regime with $N_{HI} < 10^{18}$ cm$^{-2}$
is a new frontier that awaits exploration in the local universe.

\medskip
\reference  Bahcall, J.N., et al 1993, ApJS, 87, 1
\reference  Bartelmann, M., \& Loeb, A. 1996, ApJ, 457, 529
\reference  Boyle, B.J., \& Di Matteo, T. 1995, MNRAS, 277, L63
\reference  Briggs, F.H. 1990, AJ, 100, 999
\reference  Briggs, F.H., \& Sorar, E. 1996, in Cold Gas at High Redshift,
 eds. M.N. Bremer et al, (Kluwer Academic Publ.), p 285
\reference  Briggs, F.H., Sorar, E., Kraan-Korteweg, R.C., \& van Driel, W.
  1996, this volume  
\reference  Charlton, J.C., Salpeter, E.E., \& Linder, S.M. 1993, ApJ, 430, L29
\reference  Chengalur, J.N., Giovanelli, R., \& Haynes, M.P. 1995, AJ, 109, 2415
\reference  Corbelli, E., Schneider, S.E., \& Salpeter, E.E. 1989, AJ, 97, 390
\reference  Ellis, R.S., Colless, M., Broadhurst, T.,
  Heyl, J., Glazebrook, K. 1996, MNRAS. 280, 235
\reference  Fall, S.M., \& Pei, Y.C. 1989, ApJ, 337, 7
\reference  Giovanelli, R., \& Haynes, M.P. 1989, Ap.J., 346, L5
\reference  Giovanelli, R., Scodeggio, M., Solanes, J.M., Haynes, M.P.,
  Arce, H,. \& Sakai, S. 1995, AJ, 109, 1451
\reference  Glazebrook, K., Ellis, R., Santiago, B., \& Griffiths, R. 1995,
  MNRAS, 275, L19
\reference  Henning, P.A. 1995, ApJ, 450, 578
\reference  Hewett, P.C., Foltz, C.B., \& Chaffee, F.H. 1993, Ap.J., 406, L43
\reference  Kennicutt, R.C., Tamblyn, P., and Congdon, C.W. 1994, ApJ, 435, 22
  (KTC)
\reference  Lanzetta, K.M., Wolfe, A.M., \& Turnshek, A.M. 1995, ApJ, 440, 435
\reference  Maloney, P. 1993, ApJ, 414, 41
\reference  Pei, Y.C., \& Fall, S.M. 1995, ApJ, 454, 69
\reference  Petitjean, P., Webb, J.K, Rauch, M., Carswell, R.F., \&
  Lanzetta, K.M. 1993, MNRAS, 262, 499
\reference  Rao, S.M., \& Briggs, F.H. 1995, ApJ, 419, 515 
\reference  Rao, S.M., Turnshek, D.A., \& Briggs, F.H. 1995, ApJ, 449, 488
\reference  Schneider, S.E. 1989, Ap.J., 343, 94
\reference  Schneider, S.E. 1996, this volume
\reference  Schmidt, M. Schneider, D.P., \& Gunn, J.E. 1994, AJ, 107, 1245
\reference  Sorar, E. 1994, Ph.D. Thesis, University of Pittsburgh
\reference  Smette, A., Surdej, J., Shaver, P.A., Foltz, C.B., Chaffee, F.H.,
  Weymann, R.J., Williams, R.E., \& Magain, P. 1992, ApJ, 389, 39
\reference  Smette, A., Claeskens, J.F., \& Surdej, J. 1995a, in 
 Astrophysical Implications of Gravitational  Lensing, Proc.
 of IAU Symposium 173, ``  eds. C.S. Kochanek \& J.N. Hewitt, (Kluwer
 Academic Publ), p 99
\reference  Smette, A., Robertson, J.G., Shaver, P.A., Reimers, D.,
  Wisotski, L., \& Koehler, T. 1995b, A\&AS, 113, 199 
\reference  Smette, A., et al, 1996 submitted to A\&A
\reference  Steidel, C.C. 1990, ApJS, 72, 1
\reference  Szomoru, A, Guhathakurta, P, van Gorkom, J.H., Knapen, J.H,
  Weinberg, D.H., Fruchter, A.S. 1994, AJ, 108, 491
\reference  Szomoru, A, van Gorkom, J.H., Gregg, M.D., \& Strauss, M.A. 
  1994, AJ, 111, 2150
\reference  Storrie-Lombardi, L.J., McMahon, R.G., Irwin, M.J., \& Hazard, C.
  1996, ApJ, 468, 121
\reference  Sunyaev, R.A. 1969, ApJ, 3, 33
\reference Taylor, C.L., Brinks, E., \& Skillman, E.D. 1993, AJ, 105, 128
\reference Taylor, C.L., Thomas, D.L., \& Skillman, E.D. 1996, preprint
\reference  Webster, R.L., Francis, P.J., Peterson, B.A., Drinkwater, M.J.,
  \& Masci, F.J. 1995, Nat, 375, 469
\reference  Wolfe, A.M., Turnshek, D.A., Smith, H.E., \& Cohen, R.D. 1986,
  ApJS, 61, 249
\reference  Zwaan, M., Sprayberry, D., \& Briggs, F.H. 1996, this
volume.
\end{document}